%% file: main.tex
\documentclass[10pt,pra,aps,twocolumn,superscriptaddress,nofootinbib,showkeys]{revtex4-1}

\usepackage{amsmath}  \usepackage{amssymb}   \usepackage{amsfonts}  \usepackage{bm}  \usepackage{bbm}      \usepackage{braket}    \usepackage{color}  \usepackage{comment}  \usepackage{dcolumn}  \usepackage{enumerate}  \usepackage{epsfig}  \usepackage{gensymb}  \usepackage{graphicx}  \usepackage{indentfirst}  \usepackage{lmodern}  \usepackage{mathrsfs}  \usepackage{mathtools}  \usepackage{psfrag}   \usepackage{pst-all}   \usepackage{soul}  
\usepackage{xcolor}
\usepackage{upgreek}
\usepackage{siunitx}
\usepackage{textcomp}
\usepackage{textgreek}
\usepackage{multirow}
\usepackage{placeins} 
\usepackage[T1]{fontenc} 
\usepackage[hidelinks]{hyperref}

\def\squared{^2}
      
\begin{document} 
\def\bibsection{\section*{\refname}} 

\title{Optically Induced Thermal Runaway in Phase-Change \texorpdfstring{VO$_2$}{VO2} Nanostructures}
\author{Ji\v{r}\'{i} Kab\'{a}t}
\affiliation
{Brno University of Technology, Central European Institute of Technology, Purky\v{n}ova 123, 612 00, Brno, Czech Republic}
\affiliation
{Brno University of Technology, Faculty of Mechanical Engineering, Institute of Physical Engineering, Technická 2, 616 69, Brno, Czech Republic}
\author{Peter Kepi\v{c}}
\affiliation
{Brno University of Technology, Central European Institute of Technology, Purky\v{n}ova 123, 612 00, Brno, Czech Republic}
\affiliation
{Brno University of Technology, Faculty of Mechanical Engineering, Institute of Physical Engineering, Technická 2, 616 69, Brno, Czech Republic}
\author{Andrea Kone\v{c}n\'{a}}
\affiliation
{Brno University of Technology, Central European Institute of Technology, Purky\v{n}ova 123, 612 00, Brno, Czech Republic}
\affiliation
{Brno University of Technology, Faculty of Mechanical Engineering, Institute of Physical Engineering, Technická 2, 616 69, Brno, Czech Republic}
\author{Filip Ligmajer}
\email{filip.ligmajer@vutbr.cz}
\affiliation
{Brno University of Technology, Central European Institute of Technology, Purky\v{n}ova 123, 612 00, Brno, Czech Republic}
\affiliation
{Brno University of Technology, Faculty of Mechanical Engineering, Institute of Physical Engineering, Technická 2, 616 69, Brno, Czech Republic}

\begin{abstract}

While thermal runaway occurs across diverse physical disciplines, its presence in subwavelength phase-change photonics remains unexplored because existing thermo-optical models break down near sharp optical transitions. Here, we introduce an iterative multiphysics framework coupling full-wave electrodynamics with heat transfer, and we discover optically induced thermal runaway in vanadium dioxide ($\text{VO}_2$) nanostructures. The runaway switching threshold depends on the illumination wavelength, ambient temperature, and underlying substrate, with external thermal biasing significantly reducing the optical intensity required for runaway ignition. Our results push forward the modeling of sharp photothermal transitions and provide essential design principles for active metasurfaces, neuromorphic photonic devices, and nanoscale thermal management.

\end{abstract}

\maketitle
\date{\today}

\begin{figure}[t!]
\includegraphics[width=0.47\textwidth]{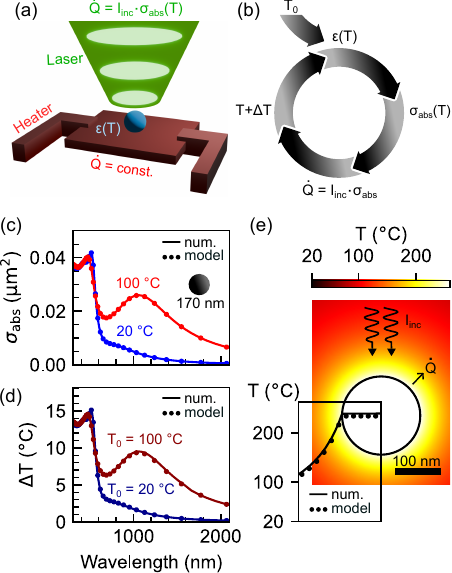}
\caption{(a) Schematic of the studied system featuring a VO$_2$ nanosphere with temperature-dependent dielectric function $\varepsilon(T)$. External heat can be provided by either a heater (absorbed power $\dot{Q}$ is constant) or by laser heating (absorbed power depends on the nanosphere's temperature). (b) Scheme of the positive-feedback loop: The nanostructure with a dielectric function $\varepsilon(T)$ at ambient temperature $T_0$ is illuminated by a monochromatic plane wave of intensity $I_\mathrm{inc}$ and absorbs a portion of the incident field depending on its absorption cross-section $\sigma_\mathrm{abs}(T)$. This absorption yields a dissipated power $\dot{Q}$ that drives a temperature increase $\Delta T$. The dielectric function is then updated to $\varepsilon(T+\Delta T)$ for the next step in the iterative loop.  (c) Calculated absorption cross-section spectra of a VO$_2$ nanosphere with a diameter of \SI{170}{\nano\metre} in air at \SI{20}{\celsius} and \SI{100}{\celsius} (below and above the phase transition temperature, respectively). (d) Temperature changes for the same VO$_2$ nanosphere and initial temperatures as in (b), depending on the illumination wavelength of a laser with intensity of~\SI{0.01}{\milli\watt/\micro\metre}$\squared$. The incident intensities are small, hence the steady-state is reached by just one iteration of the loop in (b). (e) Temperature distribution across the VO$_2$ nanosphere as in (c), illuminated by a plane wave with $I_\mathrm{inc}=$ \SI{0.25}{\milli\watt/\micro\metre\squared}. At the steady-state temperature, the absorbed power equals the power dissipated to the surroundings. The inset shows the radial dependence of the temperature. Results are obtained numerically (solid line) and by the analytical model (dots).}
\label{fig1}
\end{figure}

\begin{figure*}[t]
\includegraphics[width=1\textwidth]{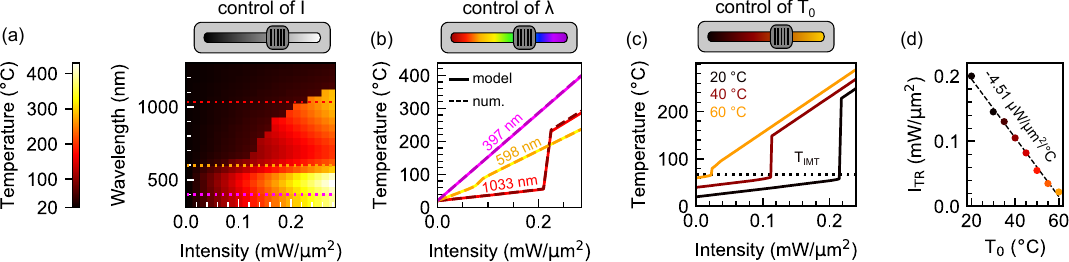}
\caption{(a) Parametric map of the steady-state temperature as a function of excitation wavelength of the laser and its intensity for the \SI{170}{\nano\metre} nanosphere. Each point was calculated using the analytical model and the iterative cycle until the temperature converged.  (b) Cross-sectional cuts of the map in (a) at three excitation wavelengths. These reveal linear heating behavior at $\lambda_\mathrm{exc}=$ \SI{397}{\nano\metre}, minor non-linearity at $\lambda_\mathrm{exc}=$ \SI{598}{\nano\metre}, and a pronounced step-like non-linearity at $\lambda_\mathrm{exc}=$ \SI{1033}{\nano\metre}. Solid lines correspond to the analytical model; dashed lines are the results of numerical simulations. (c) Effect of thermal bias introduced by the heater. Preheating the nanosphere to an ambient temperature $T_0$ reduces the illumination intensity required to ignite the thermal runaway. (d) Dependence of the thermal runaway threshold intensity $I_\mathrm{TR}$ on the preheating temperature $T_0$.}
\label{fig2}
\end{figure*}

\input{Sections/0_Introduction}

\begin{figure*}[t]
\includegraphics[width=1\textwidth]{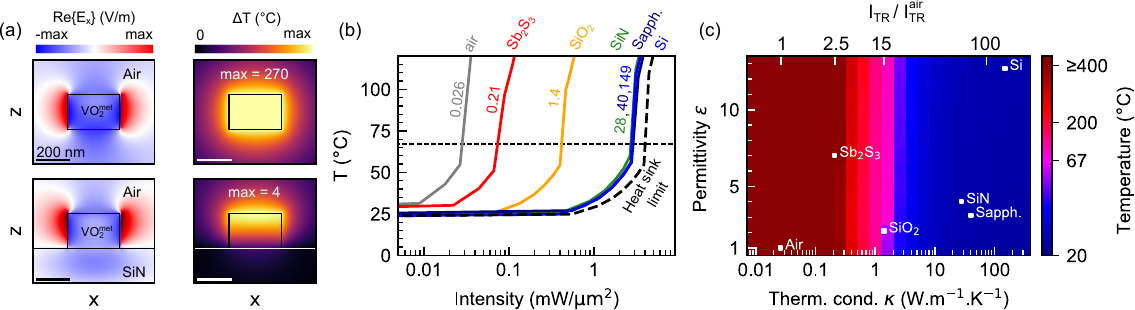}
\caption{Influence of the substrate on the photothermal heating of a metallic (met) VO$_2$ nanodisc ($D$ = \SI{300}{\nano\metre}, $h$ = \SI{200}{\nano\metre}). (a) Distribution of the induced electric field for the nanodisc evaluated for $\lambda=\SI{1033}{\nano\metre}$  (left column) and the resulting steady-state temperature change (right column). The nanodisc is either surrounded by air (first row) or placed on a SiN substrate (second row). (b) Dependence of the steady-state temperature, averaged over the nanodisc, on the illumination intensity for the nanodisc positioned on various substrates with values of $\kappa$. (c) Map of final steady-state temperature of the nanodisc as a function of substrate's permittivity $\varepsilon$ and its thermal conductivity $\kappa$. Several practically relevant substrates can be located on the map, with approximate values of incident intensity needed for the thermal runaway threshold $I_\mathrm{TR}$ compared to the one for disc located in air $I_\mathrm{TR}^\mathrm{air}$}.
\label{fig3}
\end{figure*}

\input{Sections/1_Sphere}

\begin{figure}[]
\includegraphics[width=0.47\textwidth]{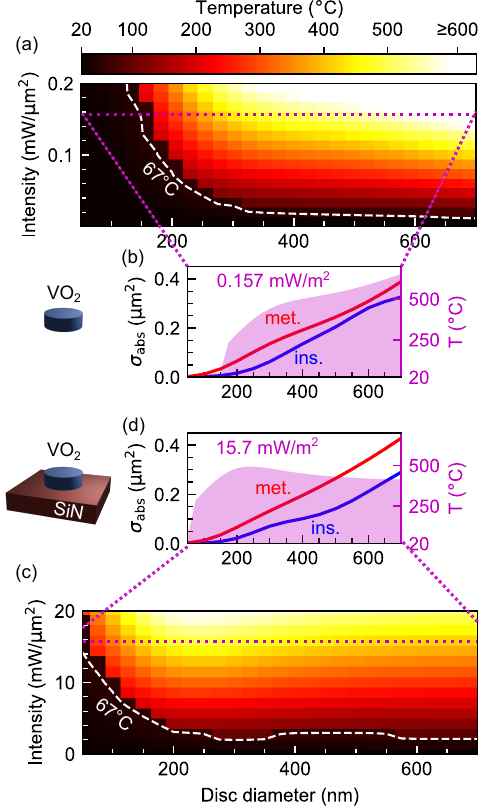}
\caption{Steady-state temperature of VO$_2$ nanodiscs. (a) Calculated intensity dependence of the steady-state temperature of the nanodiscs with different diameters $D$ and fixed height \SI{200}{\nano\metre} in air, calculated at excitation wavelength \SI{1033}{\nano\metre}. (b)  Calculated absorption cross-sections of the nanodiscs from (a) for insulating and metallic phases of VO$_2$ (increases with $D$) and the steady-state temperature (magenta) at $I_\mathrm{inc}$ (abrupt jump for nanodiscs larger than \SI{150}{\nano\metre} when IMT-boosted heating occurs). (c) Same as (a) but the nanodisc is on a SiN substrate. The intensities needed for the self-heating threshold are higher compared to those in (a).  (d) Calculated absorption cross-sections of the nanodiscs from (c). While the cross-sections for insulating and metallic VO$_2$ nanodiscs increase with $D$, the steady-state temperature (magenta) exhibits a maximum. We mark the IMT temperature \SI{67}{\celsius} in (a) and (c).}
\label{fig4}
\end{figure}

\input{Sections/2_Substrates}

\input{Sections/3_Size}

\input{Sections/Conclusions}

\begin{acknowledgments}
This work was supported by the Grant Agency of the Czech Republic (project No. 25-18336M). J.K. acknowledges the support of Brno Ph.D. Talent Scholarship -- Funded by the Brno City Municipality.
\end{acknowledgments}

\section*{Data Availability Statement}
The data supporting the findings of this study are openly available in the Zenodo repository at https://doi.org/10.5281/zenodo.22117883


\bibliographystyle{apsrev}
\bibliography{Bibliography.bib} 

\end{document}


\title{Optically Induced Thermal Runaway in Phase-Change \texorpdfstring{VO$_2$}{VO2} Nanostructures\\[2ex]
\normalfont\small\color{gray}{-- SUPPLEMENTARY INFORMATION --}}

\author{Ji\v{r}\'{i} Kab\'{a}t}
\affiliation{Brno University of Technology, Central European Institute of Technology, Purky\v{n}ova 123, 612 00, Brno, Czech Republic}
\affiliation{Brno University of Technology, Faculty of Mechanical Engineering, Institute of Physical Engineering, Technická 2, 616 69, Brno, Czech Republic}

\author{Peter Kepi\v{c}}
\affiliation{Brno University of Technology, Central European Institute of Technology, Purky\v{n}ova 123, 612 00, Brno, Czech Republic}
\affiliation{Brno University of Technology, Faculty of Mechanical Engineering, Institute of Physical Engineering, Technická 2, 616 69, Brno, Czech Republic}

\author{Andrea Kone\v{c}n\'{a}}
\affiliation{Brno University of Technology, Central European Institute of Technology, Purky\v{n}ova 123, 612 00, Brno, Czech Republic}
\affiliation{Brno University of Technology, Faculty of Mechanical Engineering, Institute of Physical Engineering, Technická 2, 616 69, Brno, Czech Republic}

\author{Filip Ligmajer}
\email{filip.ligmajer@vutbr.cz}
\affiliation{Brno University of Technology, Central European Institute of Technology, Purky\v{n}ova 123, 612 00, Brno, Czech Republic}
\affiliation{Brno University of Technology, Faculty of Mechanical Engineering, Institute of Physical Engineering, Technická 2, 616 69, Brno, Czech Republic}

\date{\today}

\maketitle

\section{Analytical calculations}
For the analytical calculations of the optical cross-sections of a spherical nanoparticle, we utilized the formalism based on the multipole expansion of the scattered electromagnetic field~\cite{Abajo1999}. The scattered field outside the sphere can be expressed using the scattering-matrix elements (analogous to the Mie coefficients $a_l$ and $b_l$~\cite{Bohren2013}), which for a homogeneous nonmagnetic sphere are given by: 
\begin{subequations}
\begin{align}
    t_l^\mathrm{E} &= \frac{-j_l(\rho_0)[\rho_1 j_l(\rho_1)]' + \epsilon [\rho_0 j_l(\rho_0)]' j_l(\rho_1)}
    {h_l^{(1)}(\rho_0)[\rho_1 j_l(\rho_1)]' - \epsilon [\rho_0 h_l^{(1)}(\rho_0)]' j_l(\rho_1)},  \\
    t_l^\mathrm{M} &= \frac{-j_l(\rho_0)\rho_1 j_l'(\rho_1) + \rho_0 j_l'(\rho_0) j_l(\rho_1)}
    {h_l^{(1)}(\rho_0)\rho_1 j_l'(\rho_1) - \rho_0 [h_l^{(1)}(\rho_0)]' j_l(\rho_1)},
\end{align}
\end{subequations}
where $\rho_0 = ka$ and $\rho_1 = k_\mathrm{in}a$. Here, $a$ is the sphere radius, $k = \omega/c$ is the free-space wave vector, $k_\mathrm{in} = \sqrt{\epsilon}\omega/c$ is the wave vector inside the sphere, $j_l(x)$ are the spherical Bessel functions of the first kind, and $h_l^{(1)}(x)$ are the spherical Hankel functions. The prime denotes differentiation with respect to $\rho_0$ or $\rho_1$.

The extinction cross-section $\sigma_\mathrm{ext}$ is then calculated as~\cite{Bohren2013}
\begin{align}
    \sigma_\mathrm{ext}=\frac{2\pi}{k\squared}\sum_{l=1}^N (2l+1)\operatorname{Re}\left( t_l^\mathrm{E} + t_l^\mathrm{M} \right),
\end{align}
and the scattering cross-section $\sigma_\mathrm{sca}$ as
\begin{align}
     \sigma_\mathrm{sca} = \frac{2\pi}{k^2} \sum_{l=1}^N (2l+1) \left( |t_l^\mathrm{E}|^2 + |t_l^\mathrm{M}|^2 \right).
\end{align}
The absorption is taken from the energy conservation $\sigma_\mathrm{abs}=\sigma_\mathrm{ext}-\sigma_\mathrm{sca}$.

For the analytical calculation of the optical heating of the nanosphere, we utilized a steady-state thermal model~\cite{Baffou2018}. Under continuous wave illumination, the nanoparticle acts as a heat source and conducts heat into the surrounding environment. Assuming the thermal conductivity of the nanoparticle is significantly higher than that of the surrounding medium, the temperature distribution inside the sphere is assumed to be uniform. Solving the heat diffusion equation yields the steady-state temperature radial profile:
\begin{align}
T(r) =T_0+\frac{\dot{Q}}{4\pi\kappa a}\quad \mathrm{for}~~r<a,
\end{align}
and outside the sphere
\begin{align}
T(r) =T_0+\frac{\dot{Q}}{4\pi\kappa r}\quad \mathrm{for}~~r>a.
\end{align}
where $T_0$ is the ambient temperature of the surrounding medium with thermal conductivity $\kappa$, and $\dot{Q}$ is the total heat power dissipated within the sphere, calculated with the incident illumination intensity $I_\mathrm{inc}$ and the absorption cross-section of the nanosphere $\sigma_\mathrm{abs}$ as $\dot{Q} = I_\mathrm{inc} \sigma_\mathrm{abs}$.

\begin{figure*}[ht!]
\includegraphics[width=1\textwidth]{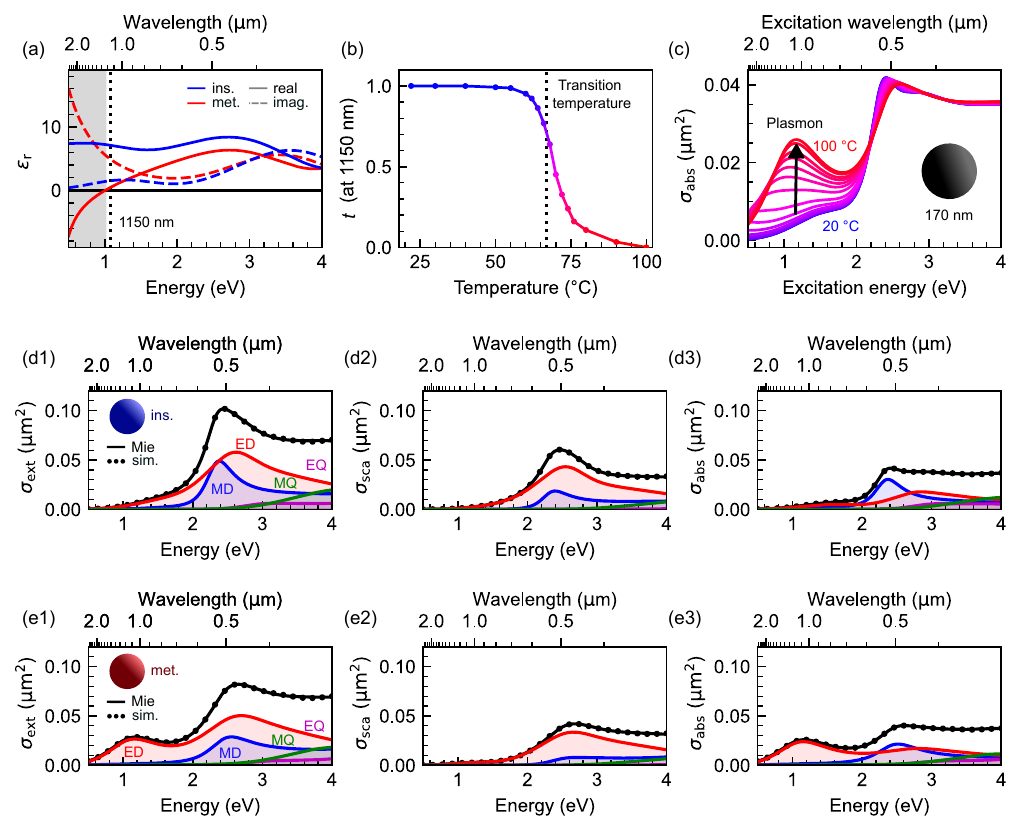}
\caption{Temperature-dependent optical properties of VO$_2$. (a) Experimentally measured dielectric function from Ref.~\cite{Kepic2025}. (b) Experimentally measured transmission of thin VO$_2$ film from Ref.~\cite{Kepic2025}. The vertical dashed line marks the phase transition temperature at \SI{67}{\celsius}. (c) Analytically calculated absorption at different temperatures. (d--e) Analytically calculated extinction (1), scattering (2), and absorption (3) cross sections of a VO$_2$ nanosphere (diameter \SI{170}{\nano\metre}) in the insulating (d) and metallic (e) phases. The dotted lines represent results obtained from numerical simulations, and the solid lines denote analytical results. We also plot the contributions of the magnetic dipole (MD, blue), electric dipole (ED, red), magnetic quadrupole (MQ, green), and electric quadrupole (EQ, magenta).
\label{SI_fig1}}
\end{figure*}

We implement the gradual temperature dependence of VO$_2$ dielectric function by weighing of temperature-dependent normalized transmission $t(T)$ measured at $\lambda_\mathrm{exc} = \SI{1150}{\nano\metre}$, plotted in Fig.~\ref{SI_fig1}(b). In the dielectric phase, $t = 1$ until around \SI{60}{\celsius} and gradually drops to $t = 0$ for the high-temperature metallic phase. We use this experimentally obtained function for weighing the experimentally measured dielectric function as
\begin{align}
    \varepsilon_\mathrm{r}(\omega,T)= t(T)\cdot\varepsilon_\mathrm{r}^\mathrm{ins}(\omega)+\left[1-t(T)\right]\cdot\varepsilon_\mathrm{r}^\mathrm{met}(\omega)
\end{align}
where $\varepsilon = \varepsilon' + i\varepsilon''$ and the superscripts stand for insulator and metal. We plot the absorption cross section for different temperatures of a VO$_2$ nanosphere in Fig.~\ref{SI_fig1}(c) and observe the gradual emergence of the plasmon peak at \SI{1}{\electronvolt}.

\section{Numerical simulations}
Numerical simulations of the optical cross-sections and thermal responses were performed using the commercial finite element software COMSOL Multiphysics, solving for both the electromagnetic and heat transfer equations.  To ensure proper convergence of the steady-state heat propagation and to prevent boundary artifacts, the finely meshed electromagnetic domain is embedded within a sufficiently large thermal domain. 

In the electromagnetic simulation, the nanostructure is excited by a background plane wave. For systems involving a substrate, the background field analytically accounts for the incident, reflected, and transmitted waves at the interface via Fresnel coefficients. The scattering cross-section of the nanostructure is evaluated by integrating the time-averaged Poynting vector of the scattered field over a surface $\partial K$ of a sphere $K$ encapsulating the structure:
\begin{align}
 \sigma_\mathrm{sca}=\frac{1}{I_\mathrm{inc}}
 \oint_{\partial \mathrm{K}}
\langle\mathbf{S}_\mathrm{sca}\rangle\cdot\hat{\mathbf{r}}\,\mathrm{d} S,\label{}
\end{align}
where $\hat{\mathbf{r}}$ is an outward pointing radial vector, $\mathrm{d} S$ a surface element, $\langle\mathbf{S}_\mathrm{sca}\rangle$ is the time-averaged Poynting vector, which is calculated as the power flow of the computed induced fields, and $I_\mathrm{inc}=\frac{1}{2}E_0^2/\sqrt{\varepsilon_0/\mu_0}$ is the intensity of the incoming field. The dissipated power density (power per unit volume) is calculated as
\begin{align}
\dot{q}_\mathrm{h} = \frac{1}{2} \mathrm{Re}\{\mathrm{i}\omega \mathbf{B} \cdot \mathbf{H}^*\} + \frac{1}{2} \mathrm{Re}\{\mathbf{J} \cdot \mathbf{E}^*\}
\end{align}
The absorption cross-section is calculated as
\begin{align}
 \sigma_\mathrm{abs}=\frac{1}{I_\mathrm{inc}}\int_\Omega \dot{q}_\mathrm{h}\,\mathrm{d}^\mathrm{3}\mathbf{r}=\frac{\dot{Q}_\mathrm{h}}{I_\mathrm{inc}} ,
\end{align}
The integration is performed over the nanostructure's volume, which renders $\dot{Q}_\mathrm{h}$ to repre the total power dissipation density inside the nanostructure.

Heat transfer simulation is performed in all domains. We solve the heat diffusion equation to obtain the temperature~$T$
\begin{align}
-\kappa\nabla\squared T = \dot{q}_\mathrm{h}
\end{align}
where $\kappa$ is the thermal conductivity, and $\dot{q}_\mathrm{h}$ is the heat power density. We then couple the radio-frequency and heat-transfer toolboxes via multiphysics coupling, with the dissipated electromagnetic power serving as a heat source. The simulation iterates until the steady-state temperature between two iterations is below \SI{0.1}{\percent} and converges. 


\newpage

\begin{table}[h]
    \centering
    \caption{Thermal properties of the materials used in the numerical simulations: density ($\rho$), thermal conductivity ($\kappa$), and specific heat ($c_\mathrm{p}$).}
    \label{tab:therm_properties}
    \begin{tabular}{c c c c c}
        Material & Ref&  $\rho$ (kg$\cdot$m$^{-3}$) & $\kappa$ (W$\cdot$m$^{-1}\cdot$K$^{-1}$) & $c_\mathrm{p}$ (J$\cdot$kg$^{-1}\cdot$K$^{-1}$) \\   
        \hline
        VO$_2$ (ins.)& \cite{Jeong2024} & 4340 & 6.5 & 690 \\
        VO$_2$ (met.)&\cite{Jeong2024} & 4340 & 6.5 & 690 \\
        Air & COMSOL & 1.2 & 0.026 & 1005 \\
        Si & \cite{Poopakdee2022} & 2329 & 149 & 713 \\
        SiO$_2$& \cite{Howes2020} & 2400 & 1.4 & 750 \\
        SiN & \cite{Karimi2025} & 3230 & 28 & 700 \\
        Sapphire & \cite{Poopakdee2022} & 3980 & 40 & 773 \\
        Sb$_2$S$_3$ (amorph.)& \cite{Aryana2023} & 4150 & 0.21 & 368 (cryst.) \\
    \end{tabular}
\end{table}

\begin{table}[b!]
    \centering
        \caption{Values of dielectric functions $\varepsilon$ of assumed environments evaluated at energy \SI{1.2}{\electronvolt} (\SI{1033}{nm}) and thermal conductivities $\kappa$.}
    \begin{tabular}{c|c|c}
        Material& $\varepsilon$ @ \SI{1.2}{\electronvolt} & $\kappa$ (W.m$^{-1}$.K$^{-1}$)\\   
        \hline
        Air& 1&0.026 \\
        Si & 12.7~\cite{Franta2017} & 149~\cite{Poopakdee2022}\\
        SiO$_2$ &2.11~\cite{Franta2016} &1.4~\cite{Howes2020}\\
        SiN & 4~\cite{Cornerstone2026} & 28~\cite{Karimi2025}\\
        Sapphire &3.1~\cite{Malitson1962} &40~\cite{Poopakdee2022}\\
        Sb$_2$S$_3$ & 7.03 & 0.21~\cite{Aryana2023}\\
    \end{tabular}
    \label{tab:Values}
\end{table}

\newpage

\newpage

\begin{figure*}[ht!]
\includegraphics[width=0.9\textwidth]{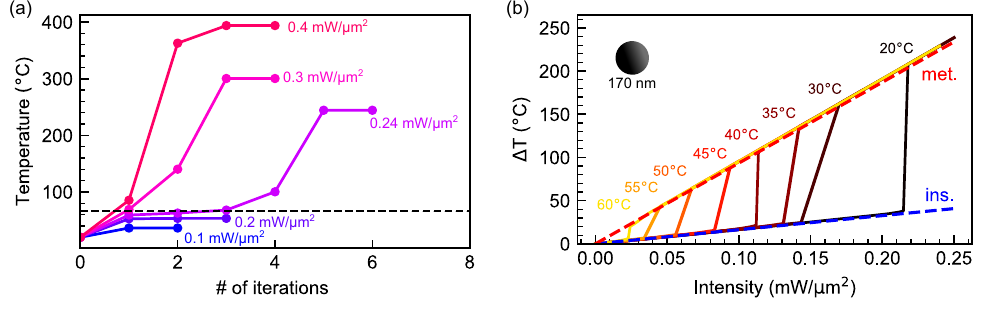}
\caption{(a) Iteratively calculated steady-state temperatures for a \SI{170}{\nano\metre} VO$_2$ hemisphere starting at an ambient temperature of $T_0 =$ \SI{20}{\celsius} for different incident intensities. (b) Effect of preheating: temperature difference $\Delta T$ for several different initial ambient temperatures $T_0$. The values are bounded by the linear heating limits of the pure insulating and metallic phases (calculated without the iterative loop). The system initially heats linearly in the insulating phase, undergoes an abrupt temperature jump upon reaching the phase transition, and subsequently resumes linear heating in the metallic phase.
}
\label{SI_fig2}
\end{figure*}
\newpage

\begin{figure*}[ht!]
\includegraphics[width=0.9\textwidth]{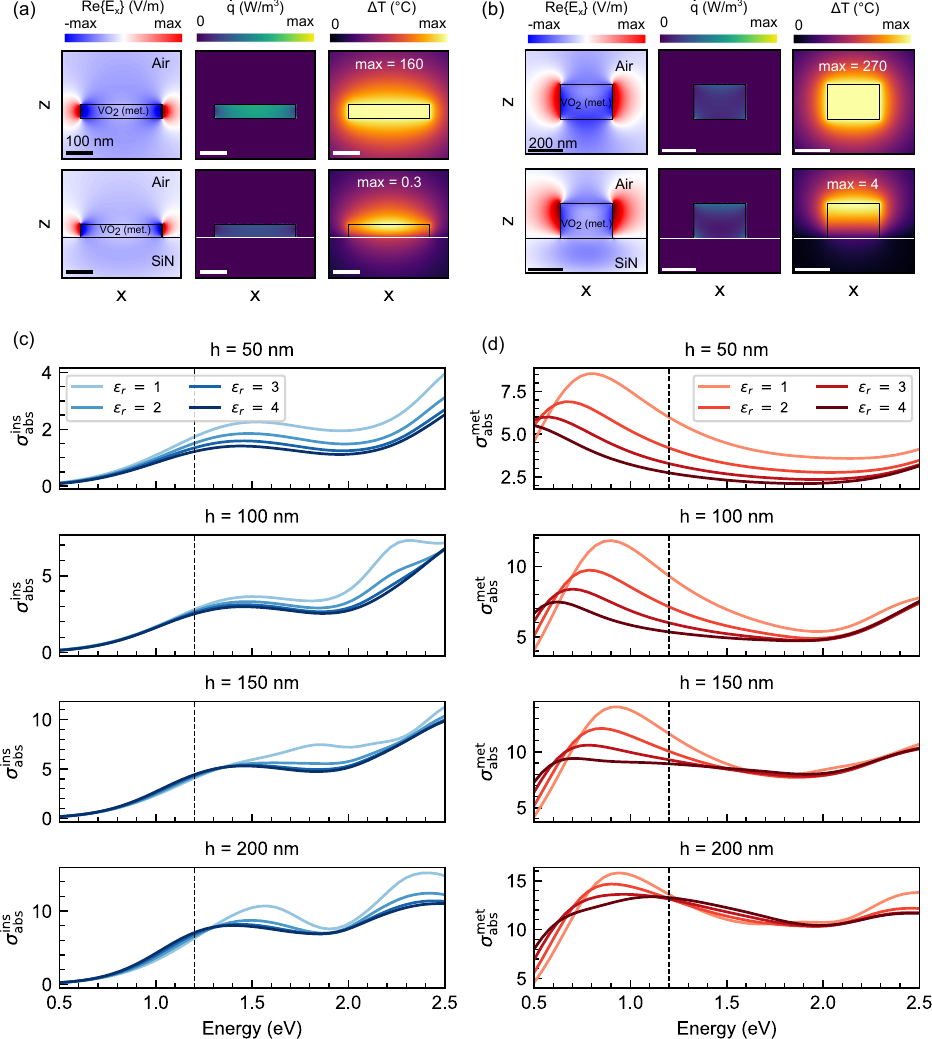}
\caption{(a) Electric field ($\mathrm{Re}\{E_x\}$), dissipated power density ($\dot{q}$), and temperature ($T$) profiles of a metallic VO$2$ nanodisc ($D =$ \SI{300}{\nano\metre}, $h =$ \SI{50}{\nano\metre}). (b) The same quantities evaluated for a thicker nanodisc ($D =$ \SI{300}{\nano\metre}, $h =$ \SI{200}{\nano\metre}). (c) Absorption cross sections $\sigma\mathrm{abs}$ of VO$_2$ nanodiscs ($D =$ \SI{300}{\nano\metre}) with various heights in the insulating (left column) and metallic (right column) phases standing on substrates with different permittivities $\varepsilon$.
\label{SI_figheight}}
\end{figure*}

\newpage
\bibliography{Bibliography.bib} 

%% file: Sections/0_Introduction.tex
Strategies for controlling light at the subwavelength scale have shifted from static resonant nanostructures to dynamically tunable metasurfaces that enable precise control over phase, amplitude, and polarization~\cite{Gu2022}. Among the various mechanisms enabling tunability~\cite{Shaltout2019}, optothermal effects are the primary control impulse for two very important classes of tunable materials: thermo-optic semiconductors~\cite{Xu2021} and phase-change materials (PCMs)~\cite{Prabhathan2023, Ya2026}, whose optical properties are naturally strongly dependent on temperature. Under optical excitation, absorbed energy induces a local temperature increase, which in turn alters the material's optical properties, creating a complex recursive feedback loop that can often lead to thermal runaway. While this effect is probably most well known in electric batteries, where the current can drive self-amplifying heat generation~\cite{Sallard2025}, it has also been studied in the context of optical resonators made of silicon~\cite{Holdman2022, Jaffe2023}, plasmonic metals~\cite{Un2020}, or even PCMs~\cite{Cortie2007, Bierman2018, Li2017}. 

Modeling of these photothermal effects frequently relies on simplifications that break down whenever sharp optothermal transitions are present. Existing methodologies often employ thermal models that assume a uniform temperature distribution across a nanostructure~\cite{Baffou2018, Fiedler2025} or use one-way coupling, in which electromagnetic results are fed into heat-transfer simulations without accounting for subsequent thermo-optical shifts~\cite{Kondo2026}. These approaches inherently lead to the loss of spatial information about dissipated heat power density and fail to predict the power required to trigger an abrupt, step-like transition in nanostructured PCMs into their often more absorptive phase.

In this Letter, we present a comprehensive thermo-optic study based on a new, inherently iterative multiphysics framework that couples electromagnetic heating and steady-state heat transfer between nanostructures and their surroundings. Crucially, this multiphysics framework enables us to precisely predict the strong optothermal nonlinearities inherent to nanostructures made of PCMs. Beyond capturing the dramatic thermal runaway associated with their phase transition, our model importantly provides complete insight into its spatial dependence. We study in detail the influence of various parameters on these strong nonlinearities, including incident intensity, nanostructure geometry, and substrate properties. We identify that the substrate's thermal conductivity and dielectric function are key factors in determining the nanostructure's final temperature. Our findings provide general guidelines for the design of building blocks for active nanophotonic devices, thermally tunable metasurfaces, and nanoscale thermal management.

\textit{Thermo-optical feedback and photothermal model}—Figure~\ref{fig1} provides a general overview of the studied system and the computational methodology. As depicted in Fig.~\ref{fig1}(a), a nanostructure, characterized by its wavelength- and temperature-dependent dielectric function $\varepsilon\left(\lambda, T\right)$, is placed on a heater (e.g., an externally heated substrate), through which we can supply a constant heat power $\dot{Q}$. In addition to external heating, the nanostructure can be heated via optical absorption upon illumination by monochromatic light. The absorbed optical power depends on the light intensity $I_\mathrm{inc}$ and on the absorption cross-section of the nanostructure $\sigma_\mathrm{abs}$, which is itself thermally dependent via $\varepsilon(T)$. Consequently, the dissipated optical power is not constant but evolves dynamically with the nanostructure temperature. The stronger the temperature dependence of $\varepsilon(T)$, the more important it is to correctly account for this effect in simulations, particularly in PCMs, where very abrupt changes in the optical properties typically emerge. For example, the insulator--metal transition (IMT) of vanadium dioxide (VO$_2$) is accompanied by one of the largest thermo-optic changes of the optical properties in nature (see Fig.~S1). Moreover, its IMT temperature of \SI{67}{\celsius} is the closest to the ambient temperature among all PCMs~\cite{Kepic2021}, and VO$_2$ thus represents an ideal material to test and validate our multiphysics methodology.

To account for these thermo-optical effects~\cite{Tognazzi2022}, we employ a thermo-optical feedback model using an iterative steady-state calculation similar to Ref.~\cite{Tsoulos2020}. The workflow of this iterative cycle is depicted in Fig.~\ref{fig1}(b). In the first step, the nanostructure at ambient temperature $T_0$ is illuminated and consequently heated by dissipated optical power, which is proportional to $\sigma_\mathrm{abs}$, which in turn depends on $\varepsilon\left(T_0\right)$. The resulting steady-state temperature of the nanostructure is then $T_0+\Delta T$. Updating  $\varepsilon\left(T_0\right)$ to $\varepsilon\left(T_0+\Delta T\right)$ then results in different $\sigma_\mathrm{abs}$ and hence different $\Delta T$. We repeat this iterative cycle $n$-times until the absolute difference between the $\mathrm{n}$-th and ($\mathrm{n}-1$)-th temperatures $|\Delta T_\mathrm{n}-\Delta T_\mathrm{n-1}|$ is smaller than a chosen tolerance. An example of the $\Delta T$ development during the calculation steps is shown in Fig.~S2.   

A similar iterative cycle was semi-analytically demonstrated in Ref.~\cite{Tsoulos2020}. Here, we extend it towards a more robust 3D multiphysics model that couples electromagnetic and thermal simulations, iterating until it reaches the steady state. To validate this numerical approach, we first benchmark it against the simplest system that can also be solved analytically: the steady-state heating of a free-standing nanosphere under low incident light intensity. The $\sigma_\mathrm{abs}$ can then be used to calculate the steady-state temperature change as~\cite{Baffou2018}

\begin{align}
\Delta T =\frac{I_\mathrm{inc}(\lambda_\mathrm{exc})\,\sigma_\mathrm{abs}(\lambda_\mathrm{exc},T)}{2\pi\kappa d},
\label{Eq_deltaT}
\end{align}

where $\lambda_\mathrm{exc}$ and $I_\mathrm{inc}$ are the wavelength and intensity of the incoming light, respectively, and $d$ is the diameter of the nanosphere, which is positioned inside a medium with thermal conductivity $\kappa$. In Fig.~\ref{fig1}(c), we show the numerically and analytically calculated $\sigma_\mathrm{abs}$ of a VO$_2$ nanosphere with $d=$ \SI{170}{\nano\metre} in air (see Supplemental Material for details of the calculations). In both the low-temperature insulating phase (blue) and the high-temperature metallic phase (red), a peak in $\sigma_\mathrm{abs}$ appears at $\approx$ \SI{500}{\nano\metre}. Based on the multipole expansion analysis, we attribute this peak to the excitation of Mie resonances~\cite{Kepic2021} (see Fig.~S1). With increasing temperature, we also observe the emergence of a strongly absorbing plasmonic dipole mode at approximately \SI{1050}{\nano\metre} (see Fig.~S1), as VO$_2$ turns metallic at these wavelengths. This strong temperature dependence of the $\sigma_\mathrm{abs}$ of VO$_2$ nanostructures thus renders them an ideal system for studying optical nanoheating. For a given nanosphere illuminated by fixed $I_\mathrm{inc}=$~\SI{0.01}{\milli\watt/\micro\metre^2}, the dependence of $\Delta T$  on the laser wavelength should replicate the spectral lineshapes of $\sigma_\mathrm{abs}$. This is confirmed by comparing Fig.~\ref{fig1}(c) with Fig.~\ref{fig1}(d), where we used Eq.~\eqref{Eq_deltaT} to calculate $\Delta T$ of the VO$_2$ nanosphere in either the insulating or metallic phase placed in air ($\kappa\approx$~\SI{0.026}{\watt\per\meter\per\kelvin}). We use these analytical results to validate our numerical simulations using the multiphysics model (see Supplemental Material for detailed methodology), achieving a very good match with a relative error of \SI{3}{\percent}. Operating in this low-intensity (\SI{0.01}{\milli\watt/\micro\metre^2}) regime ensured that the final steady-state temperature changes were small. With higher incident light intensities, however, the nanosphere could reach temperatures at which $\varepsilon$ starts to change significantly (see Fig.~S1). These one-step calculations (absorption immediately resulting in final temperature) will then fail to provide accurate predictions, and only iterative steady-state calculations will yield meaningful results, as will be discussed in the next section.

Before exploring these non-linear effects, we shift our focus to the spatial temperature distribution, remaining within the low-intensity linear regime. We plot the temperature distribution around the nanosphere heated by a plane wave ($\lambda_\mathrm{exc}=\SI{1033}{\nano\metre}$, $I_\mathrm{inc}= \SI{0.25}{\milli\watt/\micro\metre}\squared)$ in~Fig. \ref{fig1}(e). The final steady-state temperature within the nanosphere results from optical heating and thermal conduction into the environment, as schematically depicted by the arrows. In the inset, we plot the radial dependence of the temperature inside and outside the nanosphere, with the dotted line obtained from analytical calculations and the solid line from the numerical simulations (see Supplemental Material). The temperature decreases isotropically outside the nanosphere as $\sim 1/r$, while we obtain a uniform temperature distribution inside the nanosphere. The uniformity is explained by the thermal conductivity within the nanosphere being significantly higher than that of the surrounding air ($\kappa_\mathrm{VO_2}/\kappa_\mathrm{air} = 250$). Whenever this is the case, heat will be efficiently conducted throughout the entire nanosphere. This is a typical situation in thermoplasmonics \cite{Baffou2018}, whereas for thermal runaway effects on substrates, the assumption of uniform temperature may become invalid, as will be shown in the last section of this work.

%% file: Sections/1_Sphere.tex
\textit{Thermal runaway in VO$_2$ nanospheres}—To study the thermo-optical effects leading to thermal runaway, we perform the presented analytical calculation of the iteration cycle for the same system as above, but now also for excitations by plane waves of varying $I_\mathrm{inc}$ and $\lambda_\mathrm{exc}$. The cycle is set to stop when the temperature difference between two iterations falls below \SI{0.1}{\celsius}. We started at the ambient temperature $T_0=$~\SI{20}{\celsius}, and by sweeping both the excitation intensity and wavelength, we obtained a two-dimensional map of steady-state temperatures for the VO$_2$ nanosphere shown in Fig.~\ref{fig2}(a). In the region below \SI{500}{\nano\metre}, we observed a linear heating process (see the pink curve in the extracted cross-sectional cuts in Fig.~\ref{fig2}(b)). This is classical behavior for any material with optical properties only weakly dependent on temperature, as is the case of VO$_2$ in this wavelength range (cf. Fig.~\ref{fig1}(c)). In contrast, at lower excitation wavelengths, we observe sharp nonlinear temperature jumps as laser intensity increases, arising from the emergence of a plasmon resonance after the phase transition to the metallic phase. Off resonance, the non-linearity is only minor ($\lambda_\mathrm{exc}=$ \SI{598}{\nano\metre}; the yellow curve in Fig.~\ref{fig2}(b)). However, when the nanosphere is excited with the laser tuned exactly to the resonance wavelength of the plasmon in the metallic nanosphere ($\lambda_\mathrm{exc}=$ \SI{1033}{\nano\metre}; the red curve in Fig.~\ref{fig2}(b)), we can observe a typical step-like signature of thermal runaway~\cite{Holdman2022}. Here, the rapidly increasing absorption during IMT of VO$_2$ provides positive feedback to the optical heating process, and the nanosphere heats up almost in a step-like manner. From that point onward, linear heating is observed because the absorption no longer depends on temperature. Since the assumption of uniform temperature is valid here (as discussed above), we can use this system to validate our numerical simulations that employ multiphysics coupling between the electromagnetic and heat-transfer parts (see Supplemental Material for details of the model). The numerical simulations are in excellent agreement with the analytical model, as visible in~Fig.~\ref{fig2}(b).

Another degree of freedom for tuning the thermal runaway is preheating of the nanosphere, e.g., by raising the ambient temperature $T_0$ with an external heater, as sketched in Fig.~\ref{fig1}(a). The steady-state temperature as a function of laser intensity of the preheated nanosphere is plotted in Fig.~\ref{fig2}(c). The laser wavelength is fixed at \SI{1033}{\nano\metre}; thus, the black line corresponding to the starting temperature \SI{20}{\celsius} is the same as the red line in Fig.~\ref{fig2}(b). Under modest preheating to only \SI{40}{\celsius}, the required laser intensity to reach the thermal runaway threshold $I_\mathrm{TR}$ (at $T\approx$ \SI{67}{\celsius})  is reduced by a factor of two (red line). Preheating to \SI{60}{\celsius}, on the verge of the IMT, reduces $I_\mathrm{TR}$ by almost a factor of ten (orange line). We note that the step-like behavior is less prominent for preheated nanospheres, as the required intensity is lower than that for the nanosphere initially at room temperature. We plot $I_\mathrm{TR}$ for different $T_0$ in Fig.~\ref{fig2}(d), where the decreasing trend introduced by the preheating has a slope \SI{-4.51}{\micro\watt\per\square\micro\metre\per\celsius}. With the thermal runaway demonstrated on a free-standing nanosphere, we can now move to more complex effects: the influence of the substrate and size-dependent effects.

%% file: Sections/2_Substrates.tex
\textit{Influence of the substrate.}—We will now consider a more practical and realistic case of VO$_2$ nanodiscs on a substrate. Such VO$_2$ nanodiscs have been shown to exhibit tunable plasmonic and Mie resonances ~\cite{Kepic2021} and can serve as building blocks for thermally tunable metasurfaces~\cite{Naidu2026}. The presence of the substrate, however, increases the system's physical complexity by breaking its symmetry and significantly altering its optical and thermal responses. Consequently, analytical treatments become unfeasible, and theoretical analysis requires fully numerical simulations, as already suggested above. In the following, we will investigate the steady-state temperatures of VO$_2$ nanodiscs positioned on dielectric, thermally conductive substrates. We select practically relevant substrates that are non-absorbing ($\mathrm{Im}\{\varepsilon\} = 0$) at the plasmon-matching excitation wavelength of \SI{1033}{\nano\metre} (see their list, including the dielectric function and thermal conductivity values in Table~S2 in SI). 

To provide a baseline for analyzing the role of the substrate, we first present the results for a free-standing metallic VO$_2$ nanodisc in air (diameter $D =$ \SI{300}{\nano\metre}, height $h =$ \SI{200}{\nano\metre}). In the first row of Fig.~\ref{fig3}(a), we show a 2D profile of the $x$-component of the induced electric field around the nanodisc evaluated at \SI{1033}{\nano\metre}, which exhibits a characteristic plasmonic dipole shape. From the full 3D distribution of the electromagnetic fields, the dissipated heat power $\dot{Q}$ within the nanodisc is calculated. Subsequently, the numerical simulation utilizes this 3D dataset of $\dot{Q}$ as a thermal source in the coupled thermal simulation. In the temperature profile, the nanodisc is almost uniformly heated, and heat is conducted isotropically into the surrounding air. When the nanodisc is placed on a substrate (here silicon nitride, SiN), the induced electromagnetic field naturally becomes asymmetric along the $z$-axis, as shown in the lower row of Fig.~\ref{fig3}(a). The temperature profile reflects this asymmetric field and confirms that the assumption of uniformity in the analytical solution is invalid. Moreover, the overall temperature is significantly lower because the heat is efficiently conducted into the substrate. The thermal hot-spot is localized in the upper part of the nanodisc, as the surrounding air acts as a thermal insulator compared to the highly conductive substrate (which exhibits a $\kappa$ nearly three orders of magnitude larger). To analyze the substrate effect in its generality, we studied the intensity dependence of the steady-state temperature averaged over the nanodisc for the selected substrates differing in $\varepsilon$ and $\kappa$ [see Fig.~\ref{fig3}(b)]. We observe that as the thermal conductivity increases, the excitation intensity required to trigger the thermal runaway increases accordingly. We note that for Si, SiN, and sapphire substrates, we obtained remarkably similar temperature dependencies on the incident intensity. This phenomenon can be explained by convergence to the zero spreading resistance limit~\cite{Shen2022}, in which any substrate with high-enough $\kappa$ virtually acts as a non-resistive heat sink, and the temperature gradient is then confined entirely within the nanodisc. To illustrate this, we added to Fig.~\ref{fig3}(b) the corresponding intensity-temperature curve in the heat sink limit (substrate with $\varepsilon = 1$ and $\kappa = 10^{10}$). This limiting case effectively imposes the Dirichlet boundary condition ($T=T_0$) on the interface boundary, and the curves corresponding to the substrates with increasing $\kappa$ indeed converge to this limiting case.

To completely disentangle the effect of substrate's $\kappa$ and $\varepsilon$ on the thermal runaway, we perform a parametric mapping of the average steady-state temperature for the same nanodisc, illuminated by $I=\SI{0.5}{\milli\watt/\micro\metre}\squared$ in Fig.~\ref{fig3}(c). The map reveals that $\kappa$ plays the dominant role in determining the final temperature, whereas $\varepsilon$ introduces only minor changes, mainly due to the spectral shifts of the plasmonic peaks, which dictate the optical absorption in the presence of the dielectric background~\cite{Lei2015, Movsesyan2020}. Specific coordinates of $\kappa$ and $\varepsilon$ for selected substrate materials are marked in Fig.~\ref{fig3}(c) as well (see Supplemental Material for their list in Table~S2). The top y-axis clearly indicates how much higher the incident intensity needs to be compared to that in air: $2.5\times$ higher for Sb$_2$S$_3$, $15\times$ higher for SiO$_2$ and $\approx100\times$ higher for SiN, sapphire, and Si.



%% file: Sections/3_Size.tex
\textit{Size-dependent effects on heating}—We have already analyzed IMT-boosted heating of a free-standing nanosphere and of a fixed-sized nanodisc on a substrate. Now, we will explore the geometrical tunability of the nanodisc's optical absorption. While height also influences heating behavior (as detailed in Fig.~S5 in the SI), we focus primarily on the diameter, which serves as the primary tuning parameter. To systematically explore its role, we return to the simple case of a nanodisc in air for a while and fix the nanodisc height at \SI{200}{\nano\metre} (consistent with Fig.~\ref{fig3}).

In Fig.~\ref{fig4}(a), we plot the steady-state temperature maps for different diameters and incident intensities ($\lambda_\mathrm{exc} =$ \SI{1033}{\nano\metre}). It is clear that larger diameters lead to higher steady-state temperatures and to a monotonically lowered IMT-boosted heating threshold (marked by the white dashed contour). To better visualize this effect, we extract a cut from the dataset at a fixed light intensity into Fig.~\ref{fig4}(b) (purple line, right axis). Alongside the steady-state temperature, we also plot $\sigma_\mathrm{abs}$ of insulating and metallic VO$_2$ nanodiscs (blue and red curves, left axis), showing that increasing the nanodisc diameter enhances absorption. A notable deviation in the slopes occurs for small nanodiscs, where the IMT-boosted heating threshold has not yet been reached. 

Finally, we turn our attention to the realistic and practical case of nanodiscs on a SiN substrate [see Fig.~\ref{fig4}(c)]. The thermal conductivity of SiN is high compared to air, hence the incident intensities required to achieve similar temperatures are much higher [compare the intensity axes in Fig.~\ref{fig4}(a) and (c)]. The temperature map now reveals that the temperature profiles for a fixed intensity exhibit non-monotonic behavior. We again extract a cut from the temperature dataset and plot it in Fig.~\ref{fig4}(d), where the maximum steady-state temperature does not correspond to the maximum $\sigma_\mathrm{abs}$. This more complex evolution of temperature with varying nanodisc sizes can be explained by the thermal dissipation channel provided by the substrate: As the nanodisc's diameter and contact area grow, heat conduction into the substrate increases. For larger nanodiscs, therefore, the enhanced thermal conduction into the substrate outweighs the increase in absorption, resulting in a lower steady-state temperature and smaller probability of phase transition at a given illumination intensity. This effect underscores the importance of correctly accounting for the presence of the substrate, not only in terms of thermal conductivity but also in terms of the asymmetry of optical resonances~\cite{Setoura2013,Panais2023}. We presume that these effects could be experimentally verified by scanning thermal microscopy~\cite{Martinek2026}, Raman thermometry~\cite{Baffou2021}, or by transmission electron microscopy~\cite{Tizei2025,McCauley2026}.

%% file: Sections/Conclusions.tex
\textit{Conclusions}—In summary, we have analytically and numerically implemented an opto-thermal simulation loop based on the change in absorption of nanostructures during the insulator-metal transition in VO$_2$. Using this approach, we discovered a non-linear heating behavior of VO$_2$ nanostructures, analogous to the thermal runaway effect well-known in other disciplines. This effect is inherent to the strong thermal dependence of optical properties and thus will be relevant to other phase-change materials like chalcogenide glasses, other vanadium oxides, rare-earth nickelates, or perovskites~\cite{Noskin2017}, spanning the temperature range from cryo- to pyro-temperatures. 

We also explored the role of thermal bias by preheating the nanosphere, which resulted in a significant lowering of the incident intensity required for the IMT. To get closer to realistic scenarios, we also performed simulations for nanodiscs positioned on thermally conductive dielectric substrates. We showed that the substrate introduces two dominant effects: i) a change in the local induced field and absorption cross-section due to the dielectric medium, and ii) a major channel for heat conduction provided by the substrate, which results in a strong dependence of the steady state temperature on the substrate's thermal conductivity. Both of these methods of lowering the switching threshold are especially important in the domain of neuromorphic computing, where the lower switching barrier generally leads to faster switching times.

Finally, we studied how the thermal runaway depends on the size of the nanodisc and observed another non-triviality: smaller nanodiscs on a substrate may reach a higher steady-state temperature compared to their larger counterparts, despite having a smaller absorption cross section. This situation occurs because the enhanced heat conduction through the larger contact area with the substrate outweighs the respective increase in optical absorption. It would be impossible to predict all the aforementioned findings using just an analytical approach, without the spatial resolution of numerical simulations. The hidden nanoscale-specific effects may remain to be discovered in the context of polycrystalline materials, nanoparticle clusters, or nanostructure arrays. 

Our work not only describes a novel thermal runaway phenomenon in nanostructured phase-change materials but also provides fundamental guidelines for the utilization of this effect. The experimental verifications still remain to be performed, but the effects described in this work could be exploited in size-selective crystallization of VO$_2$ for control of its IMT in memory applications ~\cite{Kepic2025,Kepic2026} or in size-selective vaporization for the fabrication of monodisperse VO$_2$ nanostructure systems \cite{Zhang2010}. In analogous suitable material platforms, it could be used for size-selective laser-induced sintering~\cite{Dexter2017} or for general control of photothermal effects~\cite{Cui2023}.